\documentstyle[epsfig]{aipproc}

\begin{document}
\title{New radio observations of \\ Circinus X-1}

\author{R.~P.~Fender}
\address{Astronomy Centre, University of Sussex, Falmer, Brighton BN1 9QH, UK}

\maketitle

\begin{abstract}
New radio observations of the radio-jet X-ray binary Circinus X-1 over
nearly an entire 16.6-day orbit are presented. The source continues to
undergo radio flaring in the phase interval 0.0 -- 0.2 and appears to
be brightening since observations in the early 1990s. The radio flux
density is well correlated with simultaneous soft X-ray monitoring
from the XTE ASM, including a secondary flare event around phases 0.6
-- 0.8 observed at both energies.
\end{abstract}

\section*{Introduction}

Circinus X-1 is a highly unusual radio-bright southern X-ray binary.
Every 16.6 days the source undergoes X-ray, infrared and radio
outbursts (e.g. Glass 1994; Haynes et al. 1978). This is interpreted
as heightened accretion during periastron passage of a compact object
in an elliptical orbit around its companion star. There are few good
spectroscopic observations due in part to the presence of two 
confusing sources within 2 arcsec of the optical counterpart
(Moneti 1992), and the nature of the companion star remains uncertain.

Cir X-1 is embedded within a synchrotron nebula which trails back
towards the nearby SNR G321.9-0.3. Synthesis mapping at 6 cm 
with ATCA has resolved jet-like structures within this nebula,
originating at the location of the binary and curving back
towards G321.9-0.3 (Stewart et al. 1993) -- this suggests the 
source may be a runaway X-ray binary with an origin in the SNR.
The radio brightness of the source at cm wavelengths declined
significantly from the late 1970s to the early 1990s, with
Haynes et al. (1978) recording flux densities in excess of 1 Jy,
while Stewart et al. (1993) only measured a few mJy.

Here we present new radio monitoring of Cir X-1, over most of an orbital
period, in 1996 July, including a comparison with simultaneous soft
X-ray monitoring with the XTE ASM.

\begin{figure}[t!]
\centerline{\epsfig{file=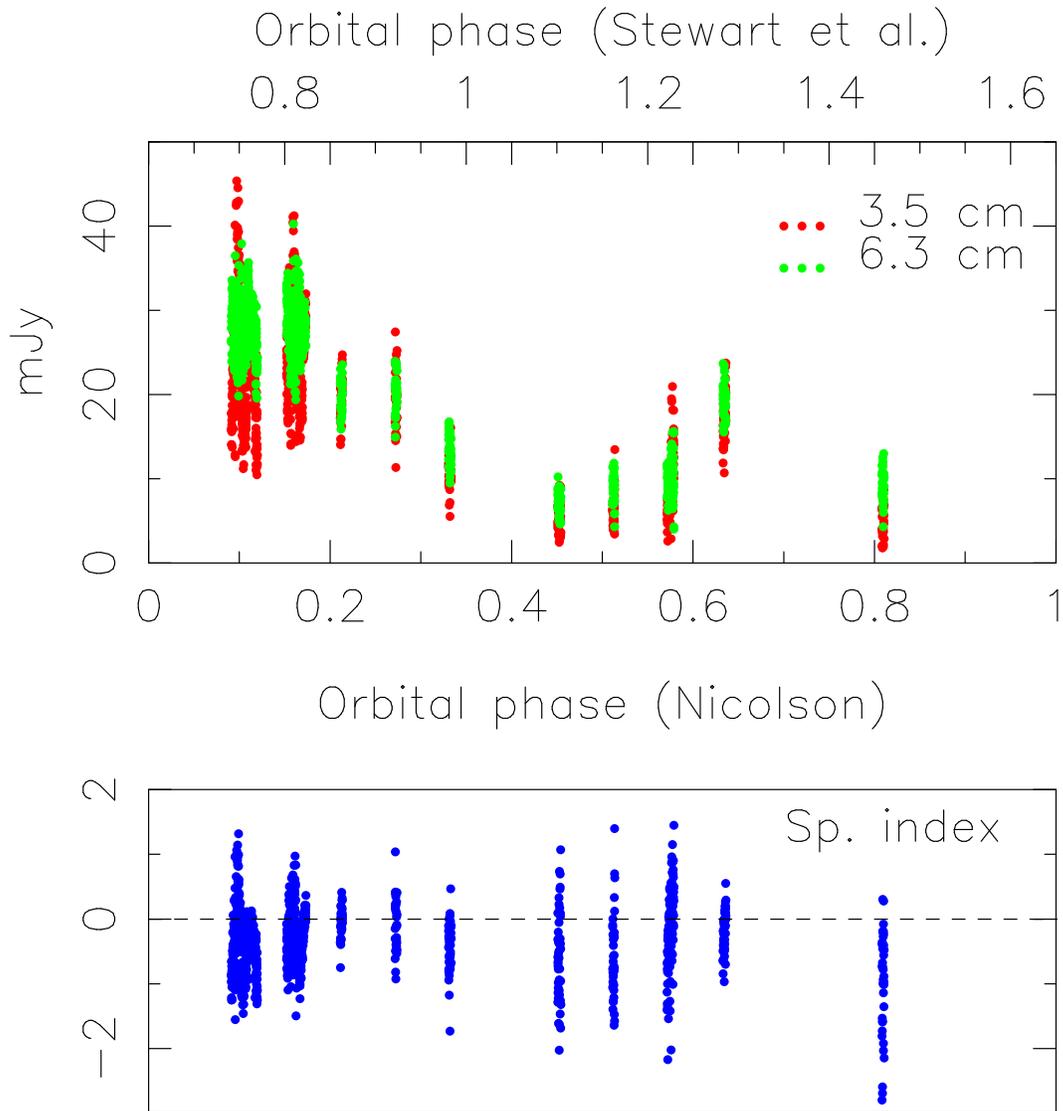,width=14cm,clip}}
\caption{ATCA observations, simultaneously at 6.3 \& 3.5 cm, of
Cir X-1 between 1996 July 1 -- 14. Orbital phase as calculated
both by the linear ephemeris of Stewart et al. and the quadratric
ephemeris of Nicolson is indicated -- clearly the Nicolson
ephemeris is more accurate. Flux density variations within a
single observation are confused by resolution and {\em u-v}
effects, but the day-to-day variations are real and the source
is clearly brightest in the phase interval 0.0 -- 0.2. The spectral
index, plotted in the lower panel, is consistent with nonthermal
synchrotron emission.}
\label{fig1}
\end{figure}

\section*{Observations}

Cir X-1 was observed with the Australia Telescope Compact Array (ATCA)
on fourteen days of its 16.6 day orbit between 1996 July 1 --
14. Observations were made simultaneously at wavelengths of 6.3 \& 3.5
cm, with the array in high resolution 6D configuration (6 km maximum
baseline). The effective resolution of the array in this configuration
is $\sim 2$ and $\sim 1$ arcsec at 6.3 \& 3.5 cm respectively.

The first two runs on Cir X-1 were of duration $\sim 12$ and $\sim 10$
hr respectively, in order to map the source at high resolution. 
However, while there is some evidence of jet-like structure in the
resultant maps at both wavelengths, uncertainty as to the contribution
both from intrinsic source variability and the surrounding nebula
render such features unreliable. The lack of much of the synchrotron
nebula in the maps suggests that it is of relatively low surface
brightness, with little structure on small (few arcsec) angular 
scales.

\section*{Results}

\subsection*{The light curve}

Fig 1 shows the radio light curve from Cir X-1 over the entire set of
observations. As described above, it is hard to differentiate
intrinsic source variability from resolution and {\em u-v} coverage
effects, and so apparent variability on short timescales should not
be taken too seriously without further careful analysis.
However, the large day-to-day changes, with a
drop in flux density by a factor of $\sim 3$ are indeed
significant. The flux density of the source, at all phases, while well
below that reported in the 1970s, appears to have risen considerably
since the observations of Stewart et al. (1993) in the early 1990s.

The spectral index ($\alpha = \Delta \log S_{\nu} / \Delta \log \nu$)
of the radio emission, plotted in the lower panel of Fig 1, is
consistent with optically thin synchrotron emission with $\alpha$
around -0.5.

\subsubsection*{A note on the ephemerides ..}

Stewart et al. (1993) discuss two ephemerides for Cir X-1 : a
quadratic ephemeris from Nicolson (private communication), and a
simplified linear ephemeris of their own. The orbital phase as
calculated from both ephemerides is shown in Figure 1; from this it is
apparent that the linear ephemeris of Stewart et al. is in error in
its prediction of the time of outburst, and the Nicolson ephemeris
must be considered more reliable. However, given that the XTE ASM has
by now observed more than thirty periodic outbursts of the system, a
newly refined X-ray ephemeris may be called for.

\subsection*{Comparison with X-ray monitoring}

Fig 2 compares the radio light curve at 6.3 cm with the 2-12 keV
lightcurve obtained by the XTE ASM. A clear correlation exists,
with flaring actvity in both energy regimes occuring between
phases 0.0 -- 0.2 followed by a subsequent decline. This supports
the picture of some enhanced accretion and related particle
acceleration/ejection occurring around the time of periastron
passage.

Note also the lesser flaring, again at {\bf both} energies, in the
phase interval 0.6 -- 0.8 (near apastron).  Such a secondary radio
outburst half an orbit after the primary flare has not previously been
reported. 

\begin{figure}[t!]
\centerline{\epsfig{file=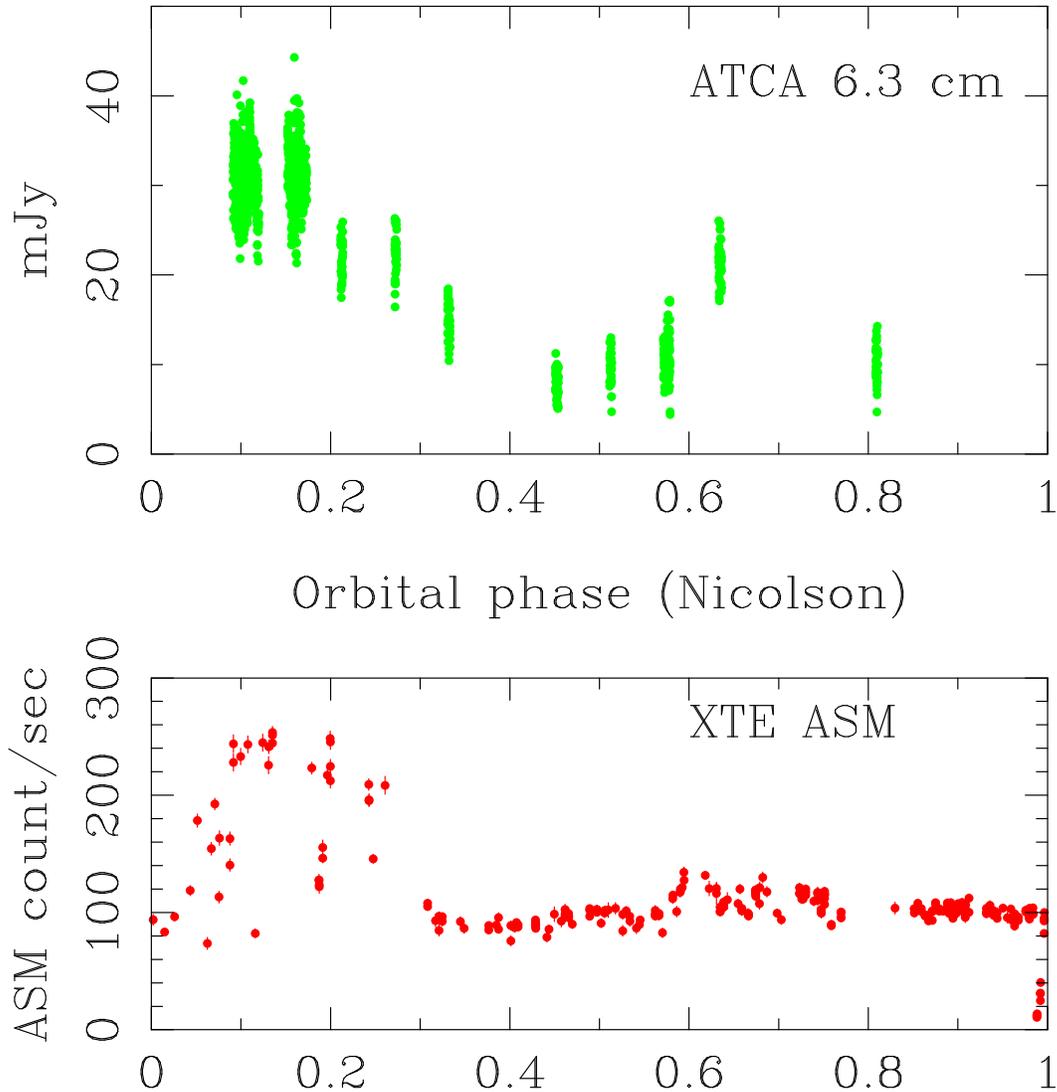,width=14cm,clip}}
\caption{A comparison of the ATCA 6.3 cm light curve with simultaneous
soft X-ray monitoring with the XTE ASM. The activity in the two
bands is clearly correlated at all phases. The major flaring,
presumably occurring around the time of periastron passage of the
compact object, is obvious. Note also however, the secondary flaring,
both in X-rays and radio emission, around phase 0.6 -- 0.8.}
\label{fig1}
\end{figure}

\section*{Conclusions}

Cir X-1 has been observed over most of its 16.6-day orbit
simultaneously at 6.3 \& 3.5 cm, and in the 2-12 keV energy range with
the XTE ASM. The source is clearly continuing to undergo radio flaring
around phase 0.0 -- 0.2 (from the quadratic ephemeris of Nicolson),
and may have begun brightening since observations in the early 1990s.
There is also a clear coupling between the soft X-ray activity
(reflecting the state of the accretion process) and the radio
brightness (probably reflecting the ejection of synchrotron-emitting
material). In addition, secondary radio flaring around phases 0.6 --
0.8, which is correlated with the X-ray behaviour, has been
discovered.

Cir X-1 is the only X-ray binary for which there is both strong
evidence for radio jets and direct evidence that the compact
object is a neutron star (from Type I X-ray bursts). It is a key
system in our understanding of radio emission, in particular
jets, from X-ray binaries. Future radio observations, including
flux monitoring, mapping and accurate proper motion measurements,
will be of great importance.

\subsection*{Acknowledgments}

I thank Jim Caswell, George Nicolson, Bob Sault, Tasso Tzioumis and
Kinwah Wu for useful discussions, and Karen Southwell and Vince
McIntyre for assistance with the observations. The ATCA is funded
by the Commonwealth of Australia for operation as a National
Facility managed by CSIRO. I acknowledge quick-look results
provided by the ASM/XTE team.

\end{document}